\documentclass[apjl,a4paper,12pt,useAMS]{emulateapj}
\usepackage{amsmath}
\usepackage{cases}
\usepackage{amssymb}
\usepackage{graphicx}
\usepackage{color}

\begin{document}
\title{Shock breakout driven by the remnant of a neutron star binary merger: An X-ray precursor of mergernova emission}
\author{Shao-Ze Li and Yun-Wei Yu}

\altaffiltext{1}{Institute of Astrophysics, Central China Normal
University, Wuhan 430079, China, {yuyw@mail.ccnu.edu.cn}}

\begin{abstract}
A supra-massive neutron star (NS) spinning extremely rapidly could
survive from a merger of NS-NS binary. The spin-down of this remnant
NS that is highly magnetized would power the isotropic merger ejecta
to produce a bright mergernova emission in ultraviolet/optical
bands. Before the mergernova, the early interaction between the NS
wind and the ejecta can drive a forward shock propagating outwards
into the ejecta. As a result, a remarkable amount of heat can be
accumulated behind the shock front and the final escaping of this
heat can produce a shock breakout emission. We describe the dynamics
and thermal emission of this shock with a semi-analytical model. It
is found that sharp and luminous breakout emission appears mainly in
soft X-rays with a luminosity of $\sim10^{45}~\rm erg~s^{-1}$ at a
few hours after the merger, by leading the mergernova emission as a
precursor. Therefore, detection of such an X-ray precursor would
provide a smoking-gun evidence for identifying NS-powered
mergernovae and distinguishing them from the radioactive-powered
ones (i.e., kilonovae or macronovae). The discovery of NS-powered
mergernovae would finally help to confirm the gravitational wave
signals due to the mergers and the existence of supra-massive NSs.
\end{abstract}

\section{Introduction}
A merger of binary neutron stars (NSs) is one of the most promising
targets of direct detection of gravitational waves (GWs). Such a
detection pointing to a few hundred Mpc is expected to come true in
the very near future with the second generation of ground-based GW
detectors (Abadie et al. 2010; Nissanke et al. 2013). For achieving
such an expectation, some observations of electromagnetic
counterparts of the GW signals are necessarily needed, which can
help to confirm the position, time, redshift, and astrophysical
properties of the GW sources.

During a NS-NS merger event, the rapidly rotating compact system
could eject a highly collimated relativistic jet and a
quasi-isotropic sub-relativistic outflow, where various
electromagnetic emission could be produced. Specifically, internal and
external dissipations of the jet energy can result in
a short-duration gamma-ray burst (GRB) and its multi-wavelength
afterglow emission, which could be the most attractive
electromagnetic counterparts in view of their high brightness (see
Nakar 2007, Berger 2014 for reviews). However,
unfortunately, the detection probability of an associated GRB is
significantly suppressed by the small opening angle of the jet (Fong
et al. 2014), although a late and weak orphan afterglow could still
be observed by off-axis (van Eerten \& MacFadyen 2011;
Metezger \& Berger 2012). Alternatively, more and more attentions
have been paid to the emission originating from the isotropic ejecta,
e.g., thermal emission due to the diffusion of internal energy of
the ejecta (called mergernova by Yu et al.
2013; Li \& Paczynski 1998; Kulkarni 2005; Rosswog 2005;
Metzger et al. 2010) and non-thermal emission due to the shock interaction
between the ejecta and environment medium (Nakar \& Piran 2011;
Metzger \& Berger 2012; Piran et al. 2013; Gao et al. 2013).

The isotropic merger ejecta is probably highly neutron-rich, which
makes it possible to effectively synthesize nuclei much heavier than
$^{56}$Ni via rapid neutron capture processes (r-processes; Rosswog
et al. 1999, 2014; Roberts et al. 2011; Goriely et al. 2011;
Korobkin et al. 2012; Bauswein et al. 2013a;  Hotokezaka et al.
2013, 2015; Wanajo et al. 2014; Goriely 2015; Just et al. 2015;
Martin et al. 2015). The radioactive decays of these newly
synthesized elements can heat the ejecta to produce a detectable
thermal emission. However, being limited by the low mass of the
ejecta (no more than a few percent of solar mass) and the element
synthesis efficiency, the luminosity of a radioactive-powered
mergernova is expected to be not much higher than $\sim10^{41}\rm
erg~s^{-1}$. Thus these phenomena are widely known as kolinova or
macronova (Kulkarni 2005; Metzger et al. 2010; Barnes \& Kasen 2013;
Tanaka \& Hotokezaka 2013; Grossman et al. 2014; Kasen et al.
2015a).

Such a luminosity limit can be breached if a more powerful energy
source can be provided by the central post-merger compact object. By
invoking a remnant supra-massive NS that spins extremely rapidly and
is highly magnetized, Yu et al. (2013) and Metzger \& Piro (2014)
investigated the characteristics of mergernova emission with an
energy injection from the spin-down of the NS. It was found that the
peak luminosity of such a NS-powered mergernova, which of course
depends on the lifetime of the NS and the collimation of the NS
wind, could sometimes be comparable to and even exceed the
luminosity of ordinary supernovae, but with a much shorter duration
on the order of a few days. Excitingly, some unusual rapidly
evolving and luminous transients were discovered by some recent
observations such as the Pan-STARRS1 Medium Deep Survey (Drout et
al. 2014). The characteristics of these transients are basically
consistent with the predictions for NS-powered mergernovae (Yu et
al. 2015), although multi-wavelength cross-identifications are still
demanded.

Recently some multi-wavelength studies have already been carried out
on observational candidates of mergernovae. For example, being
indicated by the shallow-decay afterglows of GRB 130603B, the
widely-discussed infrared excess after this GRB is argued to be
probably powered by a remnant NS (Fan et al. 2013) rather than by
radioactivities as usually considered (Tanvir et al. 2013; Berger et
al. 2013). More directly, by considering of the high-energy emission
from a NS wind (Yu et al. 2010; Zhang 2013), which can partly leak
from a preceding merger ejecta at late time, a NS-powered mergernova
is expected to be possibly accompanied by a late X-ray
re-brightening. Very recently Gao et al. (2015) found that the late
optical and X-ray bumps after GRB 080503 provided a perfect sample
for such multi-wavelength features.  In this paper, going ahead, we
would reveal another possible X-ray signature prior to a NS-powered
mergernova, which is caused by the breakout of the shock arising
from the early interaction between the NS wind and the merger
ejecta. This X-ray precursor emission would play an essential role
in future mergernova identifications and in GW detections.

\section{The model}
\subsection{Remnant neutron star and merger ejecta}

A merger would happen inevitably in a NS-NS binary when the gravity
between the NSs can no longer be supported by angular momentum
because of GW radiation release. After the merger, in some
situations, a supra-massive NS rather than a black hole could be
left with an extremely rapid differential rotation. This is
supported indirectly by many emission features of short GRBs and
afterglows including extended gamma-ray emission, X-ray flares, and
plateaus  (Dai et al. 2006; Fan \& Xu 2006; Zhang 2013; Giacomazzo
\& Perna 2013; Rowlinson et al. 2010, 2013). The presence of such a
supra-massive NS is permitted by both theoretical simulations
(Bauswein et al. 2013b; Hotokezaka et al. 2011) and observational
constraints (e.g., the present lower limit of the maximum mass of
Galactic NSs is precisely set by PSR J0348+0432 to
$2.01\pm0.04M_{\odot}$; Antoniadis et al. 2013). During the first
few seconds after the birth of a remnant NS, on one hand, a great
amount of neutrinos can be emitted from the very hot NS. On the
other hand, differential rotation of the NS can generate multipolar
magnetic fields through some dynamo mechanisms (Duncan \& Thompson
1992; Price \& Rosswog 2006; Cheng \& Yu 2014), making the NS to be
a magnetar. Consequently, during the very early stage, the neutrino
emission and ultra-strong magnetic fields together could drive a
continuous baryon outflow (Dessart et al. 2009; Siegel et al. 2014;
Siegel \& Ciolfi 2015), which provides an important contribution to
form an isotropic merger ejecta. A few of seconds later, the NS
eventually enters into an uniform rotation stage and meanwhile a
stable dipole structure of magnetic fields can form. From then on,
the NS starts to lose its rotational energy via a Poynting
flux-dominated wind. The spin-down luminosity carried by the wind
can be estimated with the magnetic dipole radiation formula as
$L_{\rm sd}=L_{\rm sd,i}\left(1+{t_{\rm}}/{t_{\rm md}}\right)^{-2}$,
where $L_{\rm sd,i}=10^{47}\mathcal R^{6}_{\rm s,6}\mathcal
B^{2}_{14}\mathcal P^{-4}_{\rm i,-3}\,\rm erg\ s^{-1}$, $t_{\rm
md}=2\times10^{5}\mathcal R^{-6}_{\rm s,6}\mathcal
B^{-2}_{14}\mathcal P^{2}_{\rm i,-3}\,\rm s$, and the zero point of
time $t$ is set at the beginning of the magnetic dipole radiation
which is somewhat later than the NS formation by several seconds.
Here $\mathcal R_{\rm s}$, $\mathcal B$,  and $\mathcal P_{\rm i}$
are the radius, dipolar magnetic filed strength, and initial spin
periods of the NS, respectively, and the convention $Q_{x}=Q/10^x$
is adopted in cgs units.

During a merger event, although the overwhelmingly majority of
matter falls finally into the central remnant NS, there is still a
small fraction of matter ejected outwards, e.g., the baryon wind
blown from the NS mentioned above. Besides that component, a
quasi-isotropic merger ejecta can also be contributed by a wind from
a short-lived disk surrounding the NS, by an outflow from the
colliding interface between the two progenitor NSs, and by a tidal
tail due to the gravitational and hydrodynamical interactions. The
latter two components are usually called dynamical components. These
ejecta components differ with each other in masses, electron
fractions, and entropies. It is difficult and nearly impossible to
describe precisely the specific constitutes and distributions of a
merger ejecta, which depend on the relative sizes of the two
progenitor NSs, equation of states of NS matter, and magnetic field
structures. In any case, according to the numerical simulations in
literature, on one hand, the mass of dynamical components could
range from $\sim10^{-4}M_{\odot}$ to a few times $\sim0.01M_{\odot}$
(Oechslin et al. 2007; Bauswein et al. 2013a; Hotokezaka et al.
2013; Rosswog 2013). On the other hand, in presence of a remnant
supra-massive NS, the mass of a neutrino-driven wind is found to be
at least higher than $3.5\times10^{-3}\, M_{\odot}$ (Perego et al.
2014), and the mass-loss rate due to ultrahigh magnetic fields is
about $10^{-3}-10^{-2}M_{\odot}\,\rm s^{-1}$ during the first $1-10$
s (Siegel et al. 2014). Therefore, by combining all of these
contributions (see Rosswog 2015 for a brief review), we would take
$M_{\rm ej}=0.01M_{\odot}$ as a reference value for the total mass
of an ejecta. Furthermore, we would adopt power-law density and
velocity distributions of this mass as follows (Nagakura et al.
2014):
\begin{eqnarray}
\rho_{\rm ej}(r,t)=\frac{(\delta-3)M_{\rm ej}}{4\pi r_{\rm
max}^{3}}\left[\left(\frac{r_{\rm min}}{r_{\rm
max}}\right)^{3-\delta}-1\right]^{-1}\left({r\over r_{\max}}\right)^{-\delta},\label{Eq.rhoa}
\end{eqnarray}
and
\begin{eqnarray}
{v_{\rm ej}(r,t)}={v_{\rm max}}\frac{r}{r_{\rm max}(t)},{~~\rm
for~~}r\leq r_{\max}(t),
\end{eqnarray}
where $r$ is the radius to the central NS, $v_{\rm max}$ is the
maximum velocity of the head of ejecta which is probably on the
order of $\sim 0.1c$, and the slope $\delta$ ranges from 3 to 4
according to the numerical simulation of Hotokezaka et al. (2013).
We fix $\delta=3.5$ as in Nagakura et al. (2014). The variation of
$\delta$ within a wide range in fact cannot significantly affect the
primary results of this paper except for an extremely high value
(e.g. $\delta>5$). The maximum radius of ejecta can be calculated by
$r_{\max}(t)\approx r_{\rm max,i}+v_{\max}t$, where $r_{\rm
max,i}\approx v_{\max}\Delta t$ with $\Delta t$ being the time on
which the dipolar magnetic field is stabilized. Correspondingly, the
minimum radius reads $r_{\rm min}(t)=r_{\rm min,i}+v_{\rm min}t$ and
its initial value could be determined by an escape radius as $r_{\rm
min,i}=\left({2GM_{\rm c}r_{\rm
max,i}^{2}}/{v_{\max}^{2}}\right)^{1/3}$, where $G$ is the
gravitational constant and $M_{\rm c}$ is the mass of the remnant
NS.

The huge energy released from a remnant NS (i.e. millisecond
magnetar) would eventually drive an ultra-relativistic wind mixing
Poynting flux and leptonic plasma, which catches up with the
preceding merger ejecta very quickly. On one hand, if this wind
always keeps Poynting-flux-dominated even until it collides with the
ejecta, the material at the bottom of ejecta could be heated by
absorbing low-frequency electromagnetic waves from the Poynting
component. On the other hand, more probably, some internal
dissipations (e.g. the ICMART processes; Zhang \& Yan 2011) could
take place in the NS wind to produce non-thermal high-energy
emission. Subsequently, a termination shock could be formed at the
interface between the wind and ejecta, if the wind magnetization has
become sufficiently low there (Mao et al. 2010). As a result, the
bottom of ejecta can be heated by absorbing high-energy photons from
the emitting wind region and/or by transmitting heat from the
neighbor hot termination-shock region. Additionally, even if we
arbitrarily assume an extreme situation that all of the wind energy
is completely reflected from the interface, the bottom material of
ejecta would also be heated due to adiabatic compression by the high
pressure of the wind. Therefore, in any case, the energy carried by
the wind can always be mostly injected into the bottom of ejecta and
heat the material there.

\subsection{Shock heating and emission}
When the bottom of a merger ejecta is heated by an injected NS wind,
a pressure balance can be built naturally between the wind and the
ejecta bottom. On one hand, the pressure balance can gradually
extend to larger radii through thermal diffusion, which however
happens very slowly for an extremely optical thick ejecta. On the
other hand,  the high pressure of the bottom material can lead
itself to expand adiabatically and to get a high speed. This speed
would result in the formation of a forward shock propagating
outwards into the ejecta.

By denoting the radius and speed of a shock front by $r_{\rm sh}$
and  $v_{\rm sh}$, respectively, the increase rate of the mass of
shocked ejecta can be calculated by
\begin{eqnarray}
\frac{dM_{\rm }}{dt}=4\pi r_{\rm sh}^{2}\left[v_{\rm sh}-v_{\rm
ej}(r_{\rm sh},t)\right]\rho_{\rm ej}(r_{\rm sh},t),
\end{eqnarray}
where $v_{\rm ej}(r_{\rm sh},t)$ and $\rho_{\rm ej}(r_{\rm sh},t)$
are the velocity and density of the upstream material just in front
of the shock. Obviously, we also have $dr_{\rm sh}=v_{\rm sh}dt$. As
the propagation of the shock, its bulk kinetic energy, which is
previously gained from adiabatic acceleration, can again be partly
converted into internal energy of the newly shocked material. The
rate of this shock heating effect can be written as
\begin{eqnarray}
H_{\rm sh}^{}={1\over 2}\left[v_{\rm sh}-v_{\rm ej}(r_{\rm
sh},t)\right]^2\frac{dM_{\rm}}{dt}.\label{Hsh}
\end{eqnarray}
Then the total internal energy accumulated by the shock, $U_{\rm sh}$, can be derived
from
\begin{eqnarray}
\frac{dU_{\rm sh}^{}}{dt^{}}=H_{\rm sh}^{}-{P_{\rm sh}^{}
}\frac{d (\epsilon V_{\rm}^{})}{dt^{}}-L_{\rm sh}^{}.\label{Ush}
\end{eqnarray}
Here $P_{\rm sh}^{} =U_{\rm sh}^{}/(3\epsilon V_{\rm}^{})$ is an
average pressure, $V_{\rm}^{}\sim{(4/3)\pi r_{\rm sh}^3}$ is the
volume of the whole shocked region experiencing an adiabatic
expansion, and the fraction $\epsilon$ is introduced by considering
that the shock-accumulated heat is mostly deposited at a small
volume immediately behind the shock front. The product of this
average pressure and the corresponding volume represents the cooling
effect due to adiabatic expansion. $L_{\rm sh}^{}$ is the luminosity
of shock thermal emission, which is caused by the diffusion and
escaping of the shock heat.

Approximately following a steady diffusion equation $L=(4\pi
r^2/3\kappa \rho_{\rm })(\partial u/\partial r)c$, where $\kappa$ is
opacity, $\rho$ and $u$ are densities of mass and internal energy,
respectively, we roughly estimate the luminosity of shock thermal
emission by
\begin{eqnarray}
L_{\rm sh}\approx {r_{\rm max}^2U_{\rm sh}^{}c\over\epsilon r_{\rm
sh}^3+ (r_{\rm max}^3-r_{\rm sh}^3)}\left[{1-e^{-(\epsilon\tau_{\rm
sh}+\tau_{\rm un})}\over \epsilon\tau_{\rm sh}+\tau_{\rm
un}}\right],\label{shemi}
\end{eqnarray}
where $\tau_{\rm sh}= {\kappa M_{\rm }}/{4\pi r_{\rm sh}^{2} }$ and
$\tau_{\rm un}=  \int_{r_{\rm sh}}^{r_{\max}} \kappa \rho_{\rm ej}
dr$ are the optical depths of the shocked and unshocked ejecta,
respectively. Here the former optical depth, which only influences
the decrease of the shock emission after breakout, is calculated by
considering that the most of shocked material is concentrated within
a thin shell behind the shock front (Kasen et al. 2015b). The value
of parameter $\epsilon$ can be fixed by equating the shock
luminosity during breakout to the simultaneous heating rate, because
after that moment freshly-injected shock heat can escape from the
ejecta nearly freely. The opacity of merger ejecta is predicted to
be on the order of magnitude of $\sim (10-100)~\rm cm^2 g^{-1}$,
which results from the bound-bound, bound-free, and free-free
transitions of lanthanides synthesized in the ejecta (Kasen et al.
2013). This value is much higher than the typical one of $\kappa =
0.2 ~\rm cm^2 g^{-1}$ for normal supernova ejecta. In this paper we
take $\kappa = 10~\rm cm^2 g^{-1}$. Fairly speaking, some reducing
effects on the opacity could exist, e.g., (1) the lanthanide
synthesis in the wind components of ejecta could be blocked by
neutrino irradiation from the remnant NS by enhancing electron
fractions (Metzger \& Fern\'{a}ndez 2014) and (2) the lanthanides in
the dynamical components could be ionized by the X-ray emission from
the NS wind (Metzger \& Piro 2014).

\subsection{Shock dynamics}
The temporal evolution of shock thermal emission, as presented in
Equation (\ref{shemi}), is obviously dependent on the dynamical
evolution of the shock. Due to the slow thermal transmission in the
optical thick ejecta, a significant pressure/temperature gradient
must exist in the ejecta during shock propagation, which would lead
the shock to be continuously accelerated. Therefore, the dynamical
evolution of such a radiation-mediated shock is completely different
from the internal and external shocks in GRB situations. For an
optical thin GRB ejecta, a pressure balance can be built throughout
the whole shocked region simultaneously with the shock propagation.
In that case, the shock velocity can be simply derived from shock
jump conditions (Dai 2004; Yu \& Dai 2007; Mao et al. 2010). On the
contrary, in the present optical thick case, a detailed dynamical
calculation of the shock in principle requires an elaborate
description of the energy and mass distributions of the ejecta (see
Kasen et al. 2015b for a 1D hydrodynamical simulation for a similar
process), which is however beyond the scope of an analytical model.
Nevertheless, a simplified and effective dynamical equation can
still be obtained by according to the energy conservation of the
system.

The total energy of the shocked region, can be written
as\footnote{For simplicity, in this paper we do not taken into
account the relativistic effects that were considered in Yu et al.
(2013) for some extremely light ejecta.} $E={1\over2}M_{\rm}v_{\rm
sh}^{2}+U^{}$, where $U^{}$ is the total internal energy of the
shocked region. For a radiation-mediated shock, the value of $U$
should be much higher than $U_{\rm sh}^{}$. In principle, the
concept ``shocked region" used here can generally include the NS
wind regions, because the mass of wind leptons is drastically
smaller than that of shocked ejecta and, more importantly, the
energy released from the NS is continuously distributed in both the
wind and ejecta through thermal diffusion, which makes them behaving
like a whole. By ignoring the relatively weak energy supply by
radioactivities, the variation of the total energy can be written as
\begin{eqnarray}
{dE}=(\xi L_{\rm sd}-L_{\rm e})dt+{1\over 2}v_{\rm ej}^2(r_{\rm
sh},t){dM_{\rm }},\label{conservation}
\end{eqnarray}
where $\xi$ represents the energy injection efficiency from the NS
wind, which could be much smaller than one if the NS wind is highly
anisotropic, and $L_{\rm e}$ is the total luminosity of the thermal
emission of merger ejecta. As a general expression, the specific
form of the energy injection is not taken into account. Substituting
the expression of $E$ into Equation (\ref{conservation}), we can get
the dynamical equation of the forward shock as
\begin{eqnarray}
\frac{dv_{\rm sh}}{dt}={1\over{M_{\rm}v_{\rm sh}}}\left[(\xi L_{\rm
sd}-L_{\rm e})-{1\over2}\left(v_{\rm sh}^2-v_{\rm
ej}^2\right)\frac{dM_{\rm}}{dt}- \frac{dU_{\rm}^{}}{dt^{}}\right].\label{vsh}
\end{eqnarray}
In order to clarify the expressions of $L_{\rm e}$ and
${dU_{\rm}^{}/dt^{}}$, we denote $\tilde{U}^{}=U^{}-U_{\rm sh}^{}$,
which represents the internal energy excluding the shock-accumulated
part. The evolution of this internal energy component can be given
by
\begin{eqnarray}
\frac{d\tilde{U}^{}}{dt^{}}=\xi L_{\rm
sd}^{}-\tilde{P}^{}\frac{dV_{\rm}^{}}{dt^{}}-L_{\rm mn}^{},\label{Us}
\end{eqnarray}
where the adiabatic cooling is also calculated with an average
pressure as $\tilde{P}^{}=\tilde{U}^{}/3V_{\rm}^{}$ and
\begin{eqnarray}
L_{\rm mn}\approx{\tilde{U}^{}c\over r_{\rm
max}}\left[{1-e^{-(\tau_{\rm sh}+\tau_{\rm un})}\over \tau_{\rm
sh}+\tau_{\rm un}}\right].
\end{eqnarray}
The above expression is different from Equation (\ref{shemi})
because the majority of internal energy of the shocked region is
deposited in the most inner part of the region, which is much deeper
than the shock front.  Then we have $L_{\rm e}=L_{\rm sh}+L_{\rm
mn}$. The emission component represented by $L_{\rm mn}$ actually is
the mergernova emission discussed in Yu et al. (2013).

By substituting Equations (\ref{Hsh}), (\ref{Ush}), and (\ref{Us})
into (\ref{vsh}), we can obtain another form of the dynamical
equation as
\begin{eqnarray}
\frac{dv_{\rm sh}}{dt}={1\over{M_{\rm}v_{\rm sh}}}\left({U\over 3V}\right){dV\over dt},\label{vsh2}
\end{eqnarray}
which is just the expression adopted in Kasen et al. (2015b) to
calculate the shock breakout for super-luminous supernovae. This
equation can be easily understood in the framework of adiabatic
acceleration of a ``fireball". As discussed above, what accelerating
the ejecta material is actually the local internal pressures at
different radii, which are much lower than the pressure of the NS
wind due to the significant delay of pressure transmission. The work
done by these varying pressures can be effectively estimated with an
average internal pressure of $(U/3V)$ with respect to a volume
variation of $dV=4\pi r_{\rm sh}^2v_{\rm sh}dt$. With the above
dynamical equation, the energy conservation and assignments of the
system can be well described. By considering that the internal
energy at the time $t$ is on the order of magnitude $U\sim L_{\rm
sd}t$, Equation (\ref{vsh2}) can naturally determine a kinetic
energy also on the same order of magnitude ${1\over2}Mv_{\rm
sh}^2\sim L_{\rm sd}t$.

\begin{figure}
\centering\resizebox{0.8\hsize}{!}{\includegraphics{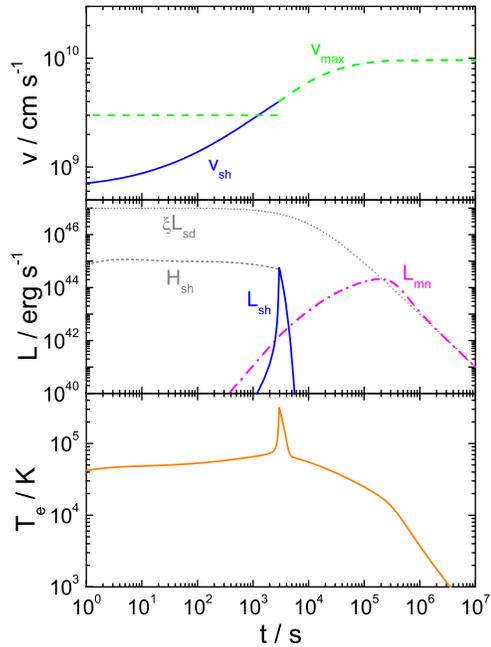}}\caption{Evolutions
of velocities (top), bolometric luminosities (middle), and emission
temperature (bottom). In the middle panel, the injected spin-down
luminosity ($\xi L_{\rm sd}$) and the shock heating rate ($H_{\rm
sh}$) are also presented for references. The model parameters are
taken as $\xi L_{\rm sd,i}=10^{47} \rm erg~s^{-1}$, $t_{\rm
md}=10^{4}$ s, $\Delta t=2$ s, $M_{\rm ej}=0.01M_{\odot}$, $v_{\rm
max}=0.1c$, $\delta=3.5$, and $\kappa=10~\rm
cm^2g^{-1}$.}\label{fitting}
\end{figure}

\section{Results and analyses}
A supra-massive NS surviving from a merger event is believed to
initially spin with a Keplerian limit period of about $1$ ms. This
corresponds to a rotational energy of several times $10^{52}$ erg with
a high stellar mass, which could be much higher in view of the rapid
differential rotation. The most of this energy is probably consumed very
quickly to generate and amplify magnetic fields, to drive a
short GRB, and maybe also to radiate GWs. The duration of this
violent stage should not be much longer than the duration of the
short GRB. So we would take $\Delta t=2$ s which is the boundary
dividing long and short GRBs. When a steady magnetic dipole
radiation begins, the spin period could have been reduced to a few
milliseconds, corresponding to an energy of $\sim10^{51-52}$ erg.
This energy supply could be further discounted for the merger ejecta
by the parameter $\xi$, if a remarkable fraction is collimated
within a small cone to power an extended gamma-ray emission and
X-ray afterglow plateau after the GRB.

\begin{figure}
\centering\resizebox{\hsize}{!}{\includegraphics{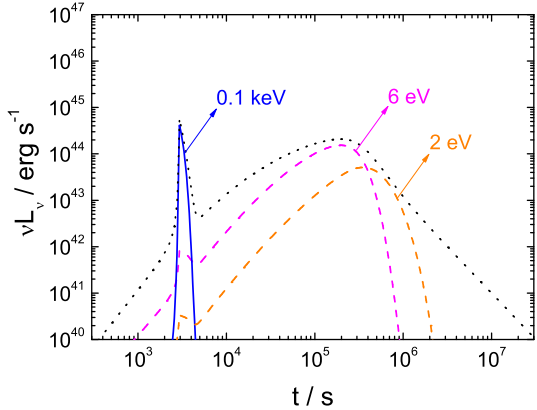}}\caption{Three
chromatic light curves for photon energies of $h \nu=0.1$ keV (soft
X-ray; solid), 6 eV (UV; dashed), and 2 eV (optical; dash-dotted),
respectively, where the bolometric light curve (dotted) is presented
for a reference.}\label{}
\end{figure}

\begin{figure}
\centering\resizebox{\hsize}{!}{\includegraphics{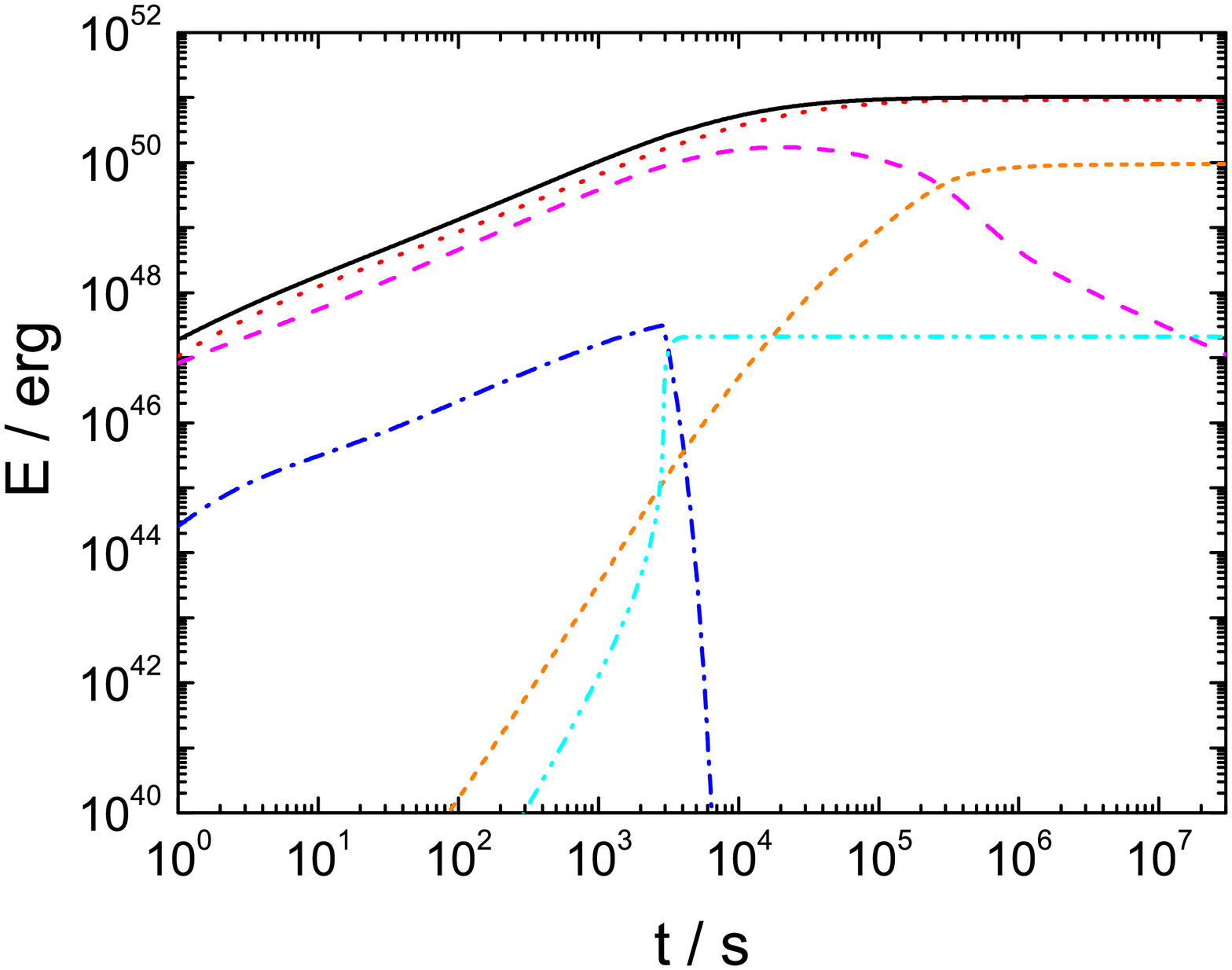}}\caption{Cumulations
of different energy components. Solid black line: total energy
provided by the NS; Dotted red line: kinetic energy of shocked
ejecta; Dashed magenta line: internal energy $\tilde{U}$;
Dash-dotted blue line: shock-produced internal energy $U_{\rm sh}$;
Dash-dot-dotted cyan line: energy of shock emission; Short-dashed
orange line: energy of mergernova emission. }\label{}
\end{figure}

For typical values of spin-down luminosity and timescale as $\xi
L_{\rm sd,i}=10^{47} \rm erg~s^{-1}$ and $t_{\rm md}=10^{4}$ s, we
present a representative numerical result in Figure 1. As shown in
the top panel, the shock initially moves slowly, with a velocity
much smaller than the maximum velocity of ejecta, and experiences a
gradual acceleration process. When the shock velocity exceeds the
maximum ejecta velocity by about a few tens percentage, shock
breakout happens, as indicated by the sharp peak in the middle
panel. It takes a remarkably long time ($\sim 10^3$ s) by the shock
to break out from the ejecta. This period is much longer than the
shock breakout time given in some previous works ($\sim1-10$ s; Gao
et al. 2013; Wang et al. 2015; Siegel \& Ciolfi 2015a,b), because
there the shock velocity was significantly overestimated by using
shock jump conditions with an assumed global pressure balance.
However, as discussed above, such a global pressure balance actually
cannot be built very quickly for a radiation-mediated shock. The
internal energy produced by the shock only occupies a very small
fraction of the total internal energy behind the shock. The shock
jump conditions could be satisfied only after a very long time
acceleration or after the ejecta is close to optical thin, before
which the shock has already crossed the whole ejecta. The middle
panel of Figure 1 also shows the shock breakout emission is very
luminous, which is comparable to that of the following bright
mergernova emission peaking at a few days. Nevertheless, since the
shock breakout and mergernova are emitted at very different radii,
the corresponding emission temperature ($\sim10^5$ K) of the former
can be much higher than the latter ($\sim10^4$ K), as presented in
the bottom panel. Here an effective temperature is defined by
$T_{\rm e}=(L_{\rm e}/4\pi r_{\rm max}^2\sigma)^{1/4}$ with $\sigma$
being the Stefan-Boltzmann constant. In more details, we further
plot three chromatic light curves in Figure 2, by assuming a
black-body spectrum\footnote{In Yu et al. (2013), this spectrum is
incorrectly written with an internal temperature rather than the
effective surface temperature used here, although the integrated
bolometric luminosity there is still right because of the reduction
by optical depth. More strictly, the surface temperature in fact
should be defined at a photosphere radius, beyond which photons can
escape freely, rather than the maximum radius. Nevertheless, before
the mergernova peak emission, the difference between these two radii
actually is always negligible (e.g., see Equation \ref{rshbo}). Thus
we approximately use the maximum radius here for simplicity. The
shift of the photosphere in the ejecta could significantly influence
the mergernova emission mainly after the peak time (Wang et al.
2015).}
\begin{eqnarray}
\nu L_{\nu}=\frac{8\pi^{2}r^{2}_{\rm
max}}{h^{3}c^{2}}\frac{(h\nu)^{4}}{{\exp}(h\nu/kT_{\rm e})-1}
\end{eqnarray}
with temperatures given in the bottom panel of Figure 1, where $h$
is the Planck constant and $k$ the Boltzmann constant. As shown,
while the mergernova emission falls into the ultraviolet band, the
shock breakout is mainly concentrated within soft X-rays, for the
adopted model parameters. Finally, in order to verify some energy
arguments mentioned above, we plot the temporal evolutions of
different energy components in Figure 3. It is exhibited that,
although the energy released from the NS is initially injected into
the ejecta in the form of internal energy, the majority of this
internal energy is finally converted into the kinetic energy of the
ejecta. As a result, during the whole optical thick period, the
internal and kinetic energies keep to be basically comparative with
each other. The internal energy produced by the shock is obviously
less than the injected one but, at the shock breakout time, the
instantaneous release of this small amount of energy can still
temporarily overshine the emission component due to the thermal
diffusion.

\begin{figure*}
\centering\resizebox{\hsize}{!}{\includegraphics{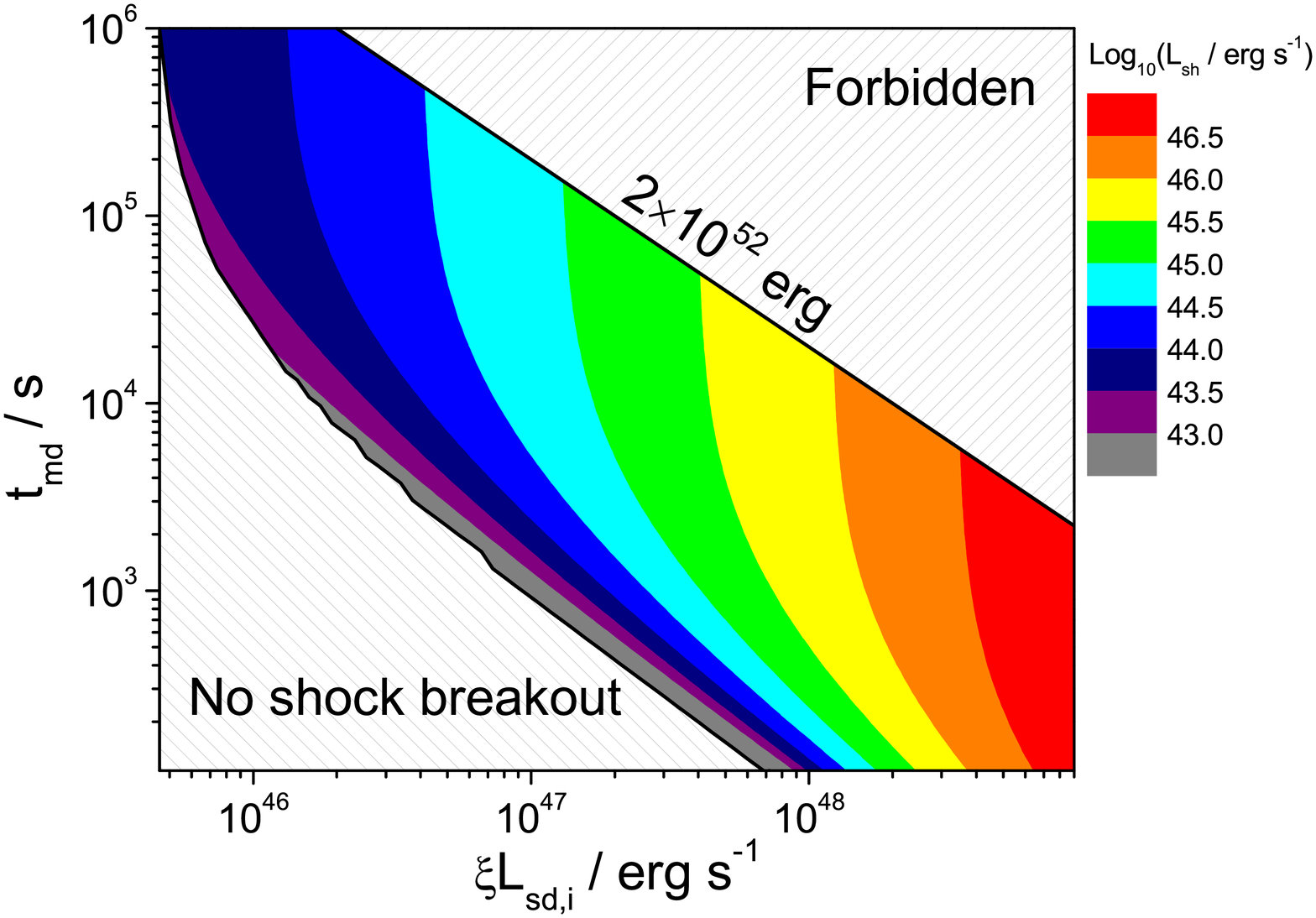}\includegraphics{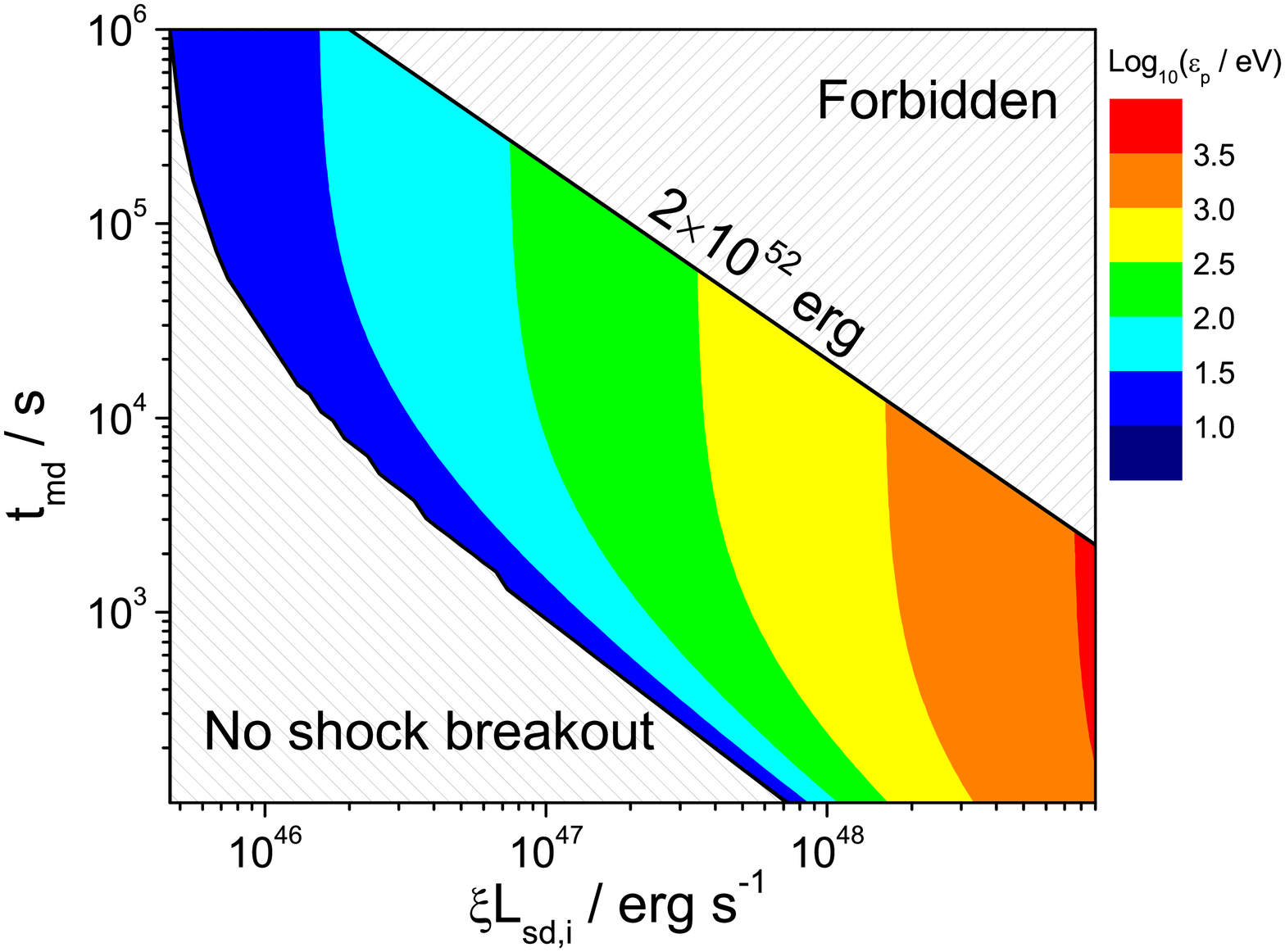}}\caption{Variations
of the luminosity (left) and peak photon energy (right) of shock
breakout emission in the $\xi L_{\rm sd,i}-t_{\rm md}$ parameter
space. The shaded region on the right-top corner is forbidden
because an unrealistically high rotational energy is required, while
in the left-bottom shaded region the shock breakout is buried in the
mergernova emission. }\label{}
\end{figure*}

For a straightforward understanding of the characteristics of the
shock breakout emission, here we provide some analytical analyses.
Firstly, in physics, shock breakout happens when the dynamical time
begins to be longer than the diffusion timescale on which photons
diffuse from the shock front to the outmost surface of merger
ejecta. Hence we can in principle solve the shock breakout time
$t_{\rm bo}$ from the equation
\begin{eqnarray}
t_{\rm bo}=t_{\rm d}&=&\left({r_{\rm max}-r_{\rm sh,bo}\over
\lambda}\right)^2{\lambda\over c}\nonumber\\
&=&{(r_{\rm max}-r_{\rm sh,bo})^2\kappa
\rho_{\rm ej}\over c},
\end{eqnarray}
where $\lambda=1/(\kappa \rho_{\rm ej})$ is the average path of
photons. Approximately we get
\begin{eqnarray}
r_{\rm sh,bo}\approx r_{\max}\left(1-{ r_{\max}\over r_{*}}\right),\label{rshbo}
\end{eqnarray}
where $r_{*}\approx(v_{\max}\kappa M_{\rm
ej}/c)^{1/2}=4.5\times10^{15}{\rm cm}$
$(v_{\max}/0.1c)^{1/2}(\kappa/10{\rm cm^{2}g^{-1}})^{1/2}(M_{\rm
ej}/0.01M_{\odot})^{1/2}$. This means, when the shock breakout
happens, the shock radius has been very close to the maximum radius
of ejecta. By invoking $U\sim\xi L_{\rm sd}t$, $P=U/(3V)$, and
acceleration rate $a\sim4\pi r_{\rm sh}^2P/M$, we can estimate the
breakout radius by $r_{\rm sh,bo}\sim{1\over2}at_{\rm
bo}^{2}\sim(\xi L_{\rm sd}/2M_{\rm ej})^{1/2}t_{\rm bo}^{3/2}$,
where $M\approx M_{\rm ej}$ is adopted because $r_{\rm sh,bo}\approx
r_{\max}$. Then, from the equation $r_{\rm sh,bo}\approx
r_{\max}=v_{\max}t_{\rm bo}$, we can simply derive the shock
breakout time to
\begin{eqnarray}
t_{\rm bo}&\sim& {2M_{\rm ej}v_{\max}^2\over\xi L_{\rm
sd}}\nonumber\\
&=&3600{\rm
s}\left({v_{\max}\over0.1c}\right)^{2}\left({M_{\rm
ej}\over0.01M_{\odot}}\right)\left({\xi L_{\rm sd}\over10^{47}\rm
erg~s^{-1}}\right)^{-1}.
\end{eqnarray}
Furthermore we get $v_{\rm sh,bo}\sim2v_{\max}$ and $r_{\rm
sh,bo}\sim{2M_{\rm ej}v_{\max}^3/\xi L_{\rm sd}}=1.1\times10^{13}\rm
cm$ that is indeed much smaller than $r_*$. Then the luminosity and
temperature of shock breakout can be estimated to
\begin{eqnarray}
L_{\rm sh,bo}&\approx&H_{\rm sh}\sim 2\pi r_{\rm sh,bo}^2(v_{\rm
sh,bo}-v_{\rm max})^3\rho_{\rm ej}\nonumber\\
&\sim&
8\times10^{45}{\rm erg~s^{-1}}\left({\xi L_{\rm sd}\over10^{47}\rm
erg~s^{-1}}\right),
\end{eqnarray}
and
\begin{eqnarray}
T_{\rm sh,bo}&=&\left({L_{\rm sh,bo}\over 4\pi r_{\rm
max}^2\sigma}\right)^{1/4}\sim6\times10^5{\rm K} \nonumber\\
&&
\times\left({v_{\max}\over0.1c}\right)^{-3/2}\left({M_{\rm
ej}\over0.01M_{\odot}}\right)^{-1/2}\left({\xi L_{\rm
sd}\over10^{47}\rm erg~s^{-1}}\right)^{3/4}.
\end{eqnarray}
The above analytical expressions qualitatively exhibits the physical
mechanisms and the parameter-dependencies of the shock breakout
emission, although the numbers given here are somewhat higher than
the numerical ones because a linear acceleration is assumed. It is
indicated that the dependence of the shock breakout emission on the
density profile of the merger ejecta (i.e., the parameter $\delta$)
is very weak. In more details, in Figure 4 we present the variations
of the shock breakout luminosity $L_{\rm sh,bo}$ and the peak photon
energy $\varepsilon_{\rm p}=4kT_{\rm sh,bo}$ in the $\xi L_{\rm
sd,i}-t_{\rm md}$ parameter space. It is indicated that, for a
significantly bight shock breakout emission, a large amount of
energy is required to be released from a NS within a sufficiently
short time.

\section{Conclusion and discussions}
Mergernovae, in particular, the ones powered by a remnant
supra-massive NS, are one of the most competitive electromagnetic
counterparts of GW signals during NS-NS mergers. The discovery of
NS-powered mergernovae could also substantially modify and expand
our conventional understandings of supernova-like transient
phenomena. Therefore, it would be an usual and essential question
that how to identify a mergernova in future searchings and
observations. Undoubtedly, a multi-wavelength method is necessary
and helpful for answering this question. In this paper, we uncover
that a shock breakout can be driven by the early interaction between
a merger ejecta and a succeeding NS wind. Such a breakout appears at
a few hours after the merger, by leading the mergernova emission as
a precursor. The breakout emission would be mainly in soft X-rays
with a luminosity of $\sim(10^{44}-10^{46})~\rm erg~s^{-1}$,
corresponding to an X-ray flux of a few$\times
(10^{-11}-10^{-9})~\rm erg~s^{-1}cm^{-2}$ for a distance of $\sim
200$ Mpc, which can be above the sensitivity of many current and
future telescopes, e.g., Swift X-ray telescope (Burrows et al.
2005), Einstein Probe (Yuan et al. 2015), etc. More optimistically,
some X-ray shock breakout emission could have been appeared in some
X-ray afterglows of short GRBs, which probably exhibits as an early
X-ray flare. It will be interesting to find out such candidates.

\acknowledgements This work is supported by the National Basic
Research Program of China (973 Program, grant 2014CB845800), the
National Natural Science Foundation of China (grant No. 11473008),
and the Program for New Century Excellent Talents in University
(grant No. NCET-13-0822).

\end{document}